# TRAINING IS EVERYTHING[1]
# ARTIFICIAL INTELLIGENCE, COPYRIGHT, AND FAIR TRAINING[2]

Dr. Andrew W. Torrance, Ph.D.
Paul E. Wilson Distinguished Professor of Law at the University of Kansas
Visiting Scholar at the MIT Sloan School of Management

Dr. Bill Tomlinson, Ph.D.
Professor of Informatics at the University of California, Irvine
Adjunct Professor, Te Herenga Waka - Victoria University of Wellington

**ABSTRACT**

Artificial intelligence ("AI") leapt into the public consciousness in 2022.[3] It did so not because of a popular Hollywood movie, like *The Terminator*, or the extravagant claim of a company or pundit. Rather, it earned this newfound attention from the public due to its sudden usefulness and practicality.[4] In quick succession, OpenAI, a software company based in San Francisco, released a graphics generator (that is, DALL-E2), a text generator (that is, GPT3.5), and then a chatbot (that is, ChatGPT) capable of carrying on compelling conversations with humans with no formal computer science training.[5] Other companies, such as Stability AI and Discord, contributed to the ready availability of AI tools easy enough for many people to use.[6] After decades of hype, AI finally achieved its first milestone of democratization.

---

[1] Mark Twain, The Tragedy of Pudd'nhead Wilson 67 (1894) ("Training is everything. The peach was once a bitter almond; cauliflower is nothing but cabbage with a college education.—*Pudd'nhead Wilson's Calendar*.").

[2] We wrote this article in collaboration with ChatGPT (Jan. 9th 2023 version). We did so, in part, to investigate how scholars and AI could collaborate to produce scholarship. While that system contributed substantially to the text, we are omitting it from the author list in line with the recommendation of Springer Nature, a major scientific publisher. *See Tools such as ChatGPT Threaten Transparent Science; Here are our Ground Rules for Their Use* Nature (Jan. 24 2023) https://www.nature.com/articles/d41586-023-00191-1.

[3] George Siemens, *Not Everything We Call an AI is Actually Artificial Intelligence. Here's what to Know*, Science Alert(Dec. 25, 2022), https://www.sciencealert.com/not-everything-we-call-an-ai-is-actually-artificial-intelligence-heres-what-to-know ("Late last month, AI, in the form of ChatGPT, broke free from the sci-fi speculations and research labs and onto the desktops and phones of the general public.").

[4] *Id.*

However, there is a *sine qua non* lurking behind these democratized sources of AI that has triggered a substantial legal response.[7] To learn how to behave, the current revolutionary generation of AIs must be trained on vast quantities of published images, written works, and sounds, many of which fall within the core subject matter of copyright law.[8] To some, the use of copyrighted works as training sets for AI is merely a transitory and non-consumptive use that does not materially interfere with owners' content or copyrights protecting it.[9] Companies that use such content to train their AI engine often believe such usage should be considered "fair use" under United States law (sometimes known as "fair dealing" in other countries).[10] By contrast, many copyright owners, as well as their supporters, consider the incorporation of copyrighted works into training sets for AI to constitute misappropriation of owners' intellectual property, and, thus, decidedly not fair use under the law.[11] This debate is vital to the future trajectory of AI and its applications.

---

[5] Johan Moreno, *OpenAI Positioned Itself as the AI Leader in 2022. But Could Google Supersede It In '23?*, Forbes (Dec. 29, 2022, 4:53 PM) https://www.forbes.com/sites/johanmoreno/2022/12/29/openai-positioned-itself-as-the-ai--leader-in-2022-but-could-google-supersede-it-in-23/?sh=13a632d55321.

[6] *See* Harry Guinness, *A Guide to the Internet's Favorite Generative AIs*, Popular Science (Jan. 11 2023, 6:00 PM), https://www.popsci.com/technology/ai-generator-guide/ (discussing the various AIs available to the public including Stability and Discord).

[7] *See* Chloe Xiang, *Artists are Suing Over Stable Diffusion Stealing Their Work for AI Art*, Vice (Jan. 17, 2023, 11:31 AM), https://www.vice.com/en/article/dy7b5y/artists-are-suing-over-stable-diffusion-stealing-their-work-for-ai-art (discussing the recently filed "class action lawsuit against Stability AI, DeviantArt, and Midjourney, alleging that the text-to-image AI tools have infringed the rights of thousands of artists and other creatives 'under the guise of 'artificial intelligence.'") and *See* Blake Brittain, *Getty Images Lawsuit says Stability AI misused photos to train AI,* Reuters (Feb. 6, 2023, 11:32 AM) https://www.reuters.com/legal/getty-images-lawsuit-says-stability-ai-misused-photos-train-ai-2023-02-06/.

[8] *See* 17 U.S.C. § 102 (1976) (copyright protection includes "literary works," "musical works," "dramatic works," "pantomimes and choreographic works," "pictorial, graphic, and sculptural works," motion pictures," "sound recordings," and "architectural works.")

[9] *See* James Vincent, *The Scary Truth About AI Copyright is Nobody Knows What Will Happen Next*, The Verge (Nov. 15, 2022, 9:00 AM) https://www.theverge.com/23444685/generative-ai-copyright-infringement-legal-fair-use-training-data (discussing the arguments in favor of the fair use defense for AI).

In this article, we analyze the arguments in favor of, and against, viewing the use of copyrighted works in training sets for AI as fair use. We call this form of fair use "fair training".[12] We identify both strong and spurious arguments on both sides of this debate. In addition, we attempt to take a broader perspective, weighing the societal costs (*e.g.*, replacement of certain forms of human employment) and benefits (*e.g.*, the possibility of novel AI-based approaches to global issues such as environmental disruption) of allowing AI to make easy use of copyrighted works as training sets to facilitate the development, improvement, adoption, and diffusion of AI. Finally, we suggest that the debate over AI and copyrighted works may be a tempest in a teapot when placed in the wider context of massive societal challenges such as poverty, equality, climate change, and loss of biodiversity, to which AI may be part of the solution.[13]

**INTRODUCTION: AI AND ITS LEAP INTO THE PUBLIC CONSCIOUSNESS**

There once was a time when the idea of talking to a machine seemed like something straight out of science fiction. Yet, in just a matter of a few short years, AI has become almost commonplace.[14] One of the pioneers of artificial intelligence, British mathematician Alan Turing, once said, "we can only see a short distance ahead, but we can see plenty there that needs to be done."[15] Today, AI has finally achieved its first true milestone of democratization, and is in the midst of changing and disrupting the way we live and work.[16]

---

[10] *See* Taysir Awad, *Universalizing Copyright Fair Use: To Copy, or Not to Copy?*, 30 J. Intell. Prop. L. 1, 3–6 (2022) (discussing the concepts of fair use and fair dealing and the countries that use each of these concepts).

[11] Vincent, *supra* note 9 (discussing the potential copyright implications of AI).

[12] We independently conceived of the phrase "fair training" ourselves. However, we do not claim we are the first to use this phrase. In fact, we would be surprised if others had not employed it previously.

[13] We have run this article through the TurnItIn plagiarism detection software to ensure that ChatGPT did not inadvertently commit plagiarism or violate copyright. As of February 28, 2023, a draft of this article had no plagiarism through TurnItIn.

[14] *See* Siemens, supra note 3.

Artificial intelligence (AI) has been a rapidly growing field in recent years, and finally gained significant usage by the general public in 2022.[17] The democratization of AI technology is largely credited to companies such as OpenAI, stability AI, and Discord, who have made it easier for individuals without formal computer science training to use and benefit from AI applications.[18] OpenAI, in particular, has released a range of AI tools including a graphics generator (DALL-E2), a text generator (GPT3.5), and a chatbot (ChatGPT) that have captured the public imagination.[19] These democratized forms of AI have paved the way for AI to be adopted in a practical mode by a wider range of people, marking a key milestone in the development of AI technology.[20]

However, with the increasing popularity of AI comes legal reactions, particularly regarding its relationship with copyright law.[21] AI algorithms must be trained on large quantities of data, such as digital images, written works, or sounds, which often fall under the purview of copyright law.[22] This raises the question of whether the unlicensed use of copyrighted works in AI training sets constitutes fair use (called "fair dealing" in some countries) under the law, or if it constitutes misappropriation of intellectual property.[23] Given the transformative nature of AI, and its potential to impact society, this issue is becoming increasingly important, relevant, and needful of resolution.

---

[15] Alan M. Turing, Computing Machinery and Intelligence, 59 Mind 433, 460 (1950).
[16] *See Generative AI Poised to Change the Way we Live According to Experts*, Virginia Tech (Jan. 31, 2023) https://vtx.vt.edu/articles/2023/01/generative-ai-experts.html.
[17] *See* Siemens, supra note 3.
[18] *See* Guinness, supra note 6.
[19] *See* Ryan Browne, *All You Need to Know about ChatGPT, the A.I. Chatbot that's Got the World Talking and Tech Giants Clashing*, CNBC (last updated Feb. 8, 2023, 10:52) https://www.cnbc.com/2023/02/08/what-is-chatgpt-viral-ai-chatbot-at-heart-of-microsoft-google-fight.html (discussing OpenAI's different AI tools).
[20] Siemens, supra note 3.
[21] *See* Xiang, supra note 7; Blake Brittain, *Getty Images Lawsuit says Stability AI misused photos to train AI,* Reuters (Feb. 6, 2023, 11:32 AM) https://www.reuters.com/legal/getty-images-lawsuit-says-stability-ai-misused-photos-train-ai-2023-02-06/.

The purpose of this article is to analyze the arguments for and against considering the unlicensed use of copyrighted works in training sets for AI as fair use, fair dealing, or "fair training".[24] The article will explore the implications of these arguments for copyright law and the future of AI technology. By examining this issue in detail, the article aims to contribute to a greater understanding of the complex relationship between AI and copyright law.

The recent rapid advance of AI marks a notable inflection point in human history. Marco Iansati and Karim Lakhani describe the singularity of this time:
> Just as in the Industrial Revolution, the age of AI is transforming the economy. However, the speed and breadth of the impact appear to be many times as great. It will not take a hundred years for digital transformation to pervade every sector of the global economy.[25]

As AI advances and becomes a more integral part of our daily lives, the need for a comprehensive examination of its legal and ethical implications becomes increasingly pressing. Given the pivotal role played by training sets, it is imperative for individuals, organizations, and policymakers to closely consider the relationship between AI and copyright law, and collaborate towards a solution that benefits society at large. This article is intended to serve as a catalyst for this much-needed discourse, and calls for a proactive approach to balancing the advancements of AI with the protections of copyright law.

---

[22] *See* Xiang, supra note 7; 17 U.S.C. § 102 (1976) (copyright protection includes "literary works," "musical works," "dramatic works," "pantomimes and choreographic works," "pictorial, graphic, and sculptural works," motion pictures," "sound recordings," and "architectural works.")

[23] Vincent, *supra* note 9 (discussing the potential copyright implications of AI).

[24] We independently conceived of the phrase "fair training" ourselves. However, we do not claim we are the first to use this phrase. In fact, we would be surprised if others had not employed it previously.

[25] *See*, Iansati, Marco, and Lakhani, Karim R. (2020), COMPETING IN THE AGE OF AI - STRATEGY AND LEADERSHIP WHEN ALGORITHMS AND NETWORKS RUN THE WORLD, page 206, Harvard Business Review Press.

**THE NEED FOR TRAINING DATA IN AI**

Artificial intelligence algorithms rely on large amounts of data to "learn" how to perform tasks and make decisions.[26] This data, referred to as "training data," is used to train AI algorithms to recognize patterns and make predictions based on those patterns.[27] The accuracy of the AI algorithm is directly dependent on the quality and quantity of the training data that it is exposed to.[28]

For example, a machine learning algorithm trained to recognize images of cats must be exposed to a large number of images of cats to learn what a cat looks like and how to distinguish it from other objects. In a similar manner, a large language model like OpenAI's GPT-3 must be exposed to large quantities of written text to learn the patterns of language and how to generate coherent and contextually appropriate responses.[29]

A problem with training data is that it often contains copyrighted works, such as images, written works, and sounds.[30] This raises the question of whether the unlicensed use of copyrighted works in training sets for AI constitutes a fair use, or fair dealing, under the law, or if it constitutes misappropriation of intellectual property.[31]

---

[26] *See generally* Amal Joby, *What is Training Data? How It's Used in Machine Learning*, Learn G2 (July 30, 2021) https://learn.g2.com/training-data (discussing the building blocks of machine learning, training data, and artificial intelligence).

[27] *Id.*

[28] *Id.*

[29] *See generally* Will Douglas Heaven, *ChatGPT is Everywhere. Here's Where it Came From*, MIT Technology Review (Feb. 8, 2023) https://www.technologyreview.com/2023/02/08/1068068/chatgpt-is-everywhere-heres-where-it-came-from/ (describing how GPT-3 functions and its capabilities).

[30] Vincent, *supra* note 9 ("Most systems are trained on huge amounts of content scraped from the web; be that text, code, or imagery.").

[31] *Id.*

Given the critical role that training data plays in AI development, it is important to understand the legal implications of using copyrighted works in AI training sets. The answer to this question has far-reaching implications for AI development and the future of AI technology. In the following sections, we will examine the arguments for and against viewing the use of copyrighted works in training sets for AI as fair use, fair dealing, or fair training.

**THE DEMOCRATIZATION OF AI THROUGH OPENAI AND OTHER COMPANIES**

Artificial intelligence was once a field that was limited to computer scientists and researchers. However, this changed dramatically in 2022 with the release of AI tools that were easy enough for people without formal computer science training to use.[32] Originally set up as a bulwark against unethical applications of AI, the company OpenAI was at the forefront of this democratization of AI, releasing graphics generators like DALL-E2, text generators like GPT-3.5, and chatbots like ChatGPT, which could carry on fluent, engaging, and sometimes even compelling conversations with humans.[33]

OpenAI's contributions to the democratization of AI were accompanied by those of several other companies, such as Stability AI and Discord, which made AI tools even more accessible to the general public.[34] With these tools, almost anyone could create and experiment with AI; from artists and musicians to journalists and small businesses, AI entered a new phase of popular accessibility.

The democratization of AI has had a profound impact on society by creating an environment in which AI is used in a more practical, everyday mode by orders of magnitude more people than ever before. There is now a new class of AI users who are

---

[32] *See* Siemens, *supra* note 3.

[33] *See* Arianna Johnson, *Here's What To Know About Open AI's ChatGPT—What It's Disrupting and How To Use It*, Forbes (Dec. 7, 2022, 12:15 P.M.) https://www.forbes.com/sites/ariannajohnson/2022/12/07/heres-what-to-know-about-openais-chatgpt-what-its-disrupting-and-how-to-use-it/?sh=47b3a0a42643.

[34] *See* Guinness *supra* note 6

not computer scientists, but rely on AI to perform a range of tasks limited largely by human imagination.[35]

Democratization of AI has also created a new set of legal challenges, as AI algorithms must be trained on vast quantities of published images, written works, and sounds, all of which are within the core subject matter of copyright.[36] The legal implications of using copyrighted works in AI training sets must be understood and addressed in order to ensure the continued growth and development of AI technology.

**THE RELATIONSHIP BETWEEN AI AND COPYRIGHT LAW**

The use of copyrighted works as training sets for AI algorithms is a new and rapidly evolving issue that has yet to be fully addressed by copyright law.[37] On one hand, some argue that the use of copyrighted works in AI training sets is a transitory and non-consumptive use that does not materially interfere with owners' copyrights, and therefore should be considered a particular form "fair use" under US law, or "fair dealing" in other countries.[38] This is "fair training".

On the other hand, others argue that the incorporation of copyrighted works into training sets for AI constitutes an unauthorized misappropriation of owners' intellectual property, and is decidedly not fair use, fair dealing, or fair training under the law[39]. This disagreement has led to conflicting interpretations of copyright law, and a lack of clarity regarding the legal status of AI training sets.[40]

---

[35] *See* Megan Cerullo, *Here's How Professionals in 3 Different Fields are Using ChatGPT for Work*, CBS News (Feb. 9, 2023, 5:00 A.M.) https://www.cbsnews.com/news/chatgpt-work-real-estate-finance-health-care-how-workers-use-it-jobs/(detailing how professionals in real estate, finance, and the medical field use ChatGPT); Nick Bilton, ChatGPT Made Me Question What It Means To Be a Creative Human, Vanity Fair (Dec. 9, 2022) https://www.vanityfair.com/news/2022/12/chatgpt-question-creative-human-robotos (describing how ChatGPT is also used to produce various forms of creative content including jokes, haikus, and screenplays).

[36] *See* Xiang, supra note 7; Brittain, *supra* note 7.

[37] *See* Vincent, supra note 9.

[38] *Id.* ("The justification used by AI researchers, startups, and multibillion-dollar tech companies alike is that using these images is covered (in the US, at least) by fair use doctrine, which aims to encourage the use of copyright-protected work to promote freedom of expression.").

It is important to consider both the legal and ethical implications of using copyrighted works in AI training sets. This includes considering the impact on copyright owners, as well as the benefits to society and the advancement of AI technology.

In order to address these questions, we must examine the current state of copyright law, as well as to consider possible solutions that may reconcile the conflicting interests of copyright owners and AI developers. This section will provide an overview of the relationship between AI and copyright law, including the legal implications of using copyrighted works in AI training sets, and the ongoing debate over the fairness of such uses.

In this article, we propose the concept of "fair training" for AI. We argue that the use of copyrighted works as training data for artificial intelligence should be considered a lawful, non-consumptive, and transformative use.[41] To understand why, it's helpful to consider how humans interact with and learn from copyrighted content.

Just like AI algorithms, humans consume, process, and store information contained within copyrighted works, such as books, music, and movies. This consumption and learning process may not infringe on the copyright of the authors in some cases, because humans have the ability to engage in transformative uses of copyrighted content.[42] For example, if a human were reading a book, she might take notes, and then summarize the book's contents, which may rise to the level of a transformative use that does not infringe on the copyright of the author.[43] Similarly, a DJ sampling tiny bits of copyrighted songs at a dance party to generate a fun musical pastiche may sometimes amount to a transformative use that does not infringe on the copyright of the original music's authors.[44]

---

[39] *See* Jessica L. Gillotte, *Copyright Infringement in Ai-Generated Artworks*, 53 U.C. Davis L. Rev. 2655, 2679–91 (2020) (discussing the circuit courts that do find infringement when AI uses copyrighted works).
[40] *Id.*
[41] *See generally* Neil Weinstock Netanel, *Making Sense of Fair Use*, 15 Lewis & Clark L. Rev. 715 (2011) (for an in-depth discussion on fair use factors).
[42] *See also* David E. Shipley, *A Transformative Use Taxonomy: Making Sense of the Transformative Use Standard*, 63 Wayne L. Rev. 267, 279–311 (2018) (defining transformative use and discussing various types of transformative use).

We argue that training AI algorithms should also be considered a transformative use and therefore, should be considered "fair training." The use of copyrighted works as training data is crucial for the development of AI, as it allows the algorithms to learn, understand, and improve upon the information they are processing. When the AI algorithm uses this training set, they are transforming the data into new and unique forms of knowledge and not producing copies of the original works. Because the AI algorithms are transforming the original work, this use should not be considered a violation of the creators' copyright. Instead, such uses by AI algorithms should be protected under a "fair training" exception.[45]

In this light, "fair training" becomes a necessary concept for the democratization and continued development of AI. The "fair training" exception will balance the rights of copyright owners with the AI's ability to learn and grow.

**ARGUMENTS IN FAVOR OF "FAIR TRAINING"**

Fair training is necessary for the continued development of AI and for society to fully realize the benefits that come from AI. AI learns in a comparable way to how humans learn, by exposure to a variety of works without necessarily violating copyright. The exposure to these sources is necessary for AI to develop the ability to recognize and understand the nuances of language, images, and sounds. The exposure ensures that AI can learn and become more sophisticated. Furthermore, fair training does not consume the data upon which it trains, but leaves this data unaltered and intact once its training is completed. Fair training is necessary for the continued development of AI and for society to fully realize the benefits that come from AI. AI learns in a comparable way to how humans learn, by exposure to a variety of works without necessarily violating copyright. The exposure to these sources is necessary for AI to develop the ability to recognize and understand the nuances of language, images, and sounds. The exposure ensures that AI can learn and become more sophisticated.

---

[43] *Id.*

[44] *Id.*

[45] *Id.* ("The use of a copyrighted work need not alter or augment the work to be transformative in nature. Rather, it can be transformative in function or purpose without altering or actually adding to the original work.") (citing A.V. v. iParadigms, LLC., 562 F.3d 630, 639 (9th Cir. 2009)).

Fair training protects the AI's ability to learn and develop because the use of copyrighted works is crucial to training the AI. The use of copyrighted works is a necessary step towards creating a more advanced AI that will benefit society. Once trained, AI transforms the copyrighted works and generates new and original works based on what it has learned, rather than copying or replicating existing works. Because the new and original works are not copies of the original, there is no commercial exploitation.

Additionally, the development of AI-powered tools and applications will lead to the creation of new jobs, the growth of existing industries, and innovative technologies.[46] These benefits will drive economic growth and benefit society in numerous ways. Moreover, the widespread adoption of AI will lead to improved efficiency and accuracy in various fields, such as healthcare, finance, and education.[47]

In summary, the arguments in favor of fair training are centered around the idea that using copyrighted works as training sets for AI is a non-consumptive, necessary, and beneficial use that promotes the advancement of AI and the growth of society. As such, it should not be considered copyright infringement.

---

[46] *See also* Adi Gaskell, *AI Creates Job Disruption Not Job Destruction*, Forbes (Jan. 18, 2022, 8:45 A.M.)
https://www.forbes.com/sites/adigaskell/2022/01/18/ai-creates-job-disruption-but-not-job-destruction/?sh=2f12ab223b3e (discussing AI's influence in the workplace)

[47] *See* Q.ai, *Artificial Intelligence's New Role in Medicine, Finance and Other Industries — How Computer Learning is Changing Every Corner of The Market*, Forbes (Feb. 2, 2023, 12:49 P.M.)
https://www.forbes.com/sites/qai/2023/02/02/artificial-intelligences-new-role-in-medicine-finance-and-other-industrieshow-computer-learning-is-changing-every-corner-of-the-market/?sh=5618c0d92e68 (discussing AI's impact in healthcare, finance, and education).

**ARGUMENTS AGAINST "FAIR TRAINING"**

Critics of the concept of "fair training" argue that using copyrighted works as training sets for AI does materially interfere with owners' copyrights and is not a transformative or non-consumptive use .[48] They view the incorporation of copyrighted works into training sets for AI as a misappropriation of owners' intellectual property and not a fair use, fair dealing, or fair training under the law.

One argument is that AI algorithms are designed to mimic human thought processes, so the use of copyrighted works in training sets may result in AI that creates similar or identical works, which would infringe on the original creators' rights.

Another argument is that the use of copyrighted works in AI training sets creates derivative works, which are protected under copyright law. This would mean that the training of AI algorithms would require permission from the copyright holders, even if the AI-generated outputs are not identical to the original works.

Additionally, critics argue that the use of copyrighted works in AI training sets could lead to market harm, as AI-generated outputs could compete with or replace the original works. The harm to the copyright holders' market can be justified as fair use, fair dealing, or fair training.

In conclusion, those who argue against "fair training" believe that the use of copyrighted works in AI training sets is an infringing use that holds the potential to harm copyright holders, and, as with derivative works, cannot be justified as fair use, fair dealing, or fair training under the law.

---

[48] *See* James Vincent, *Getty Images Sues AI Art Generator Stable Diffusion in the US for Copyright Infringement*, The Verge (Feb. 6, 2023, 10:56 A.M.) https://www.theverge.com/2023/2/6/23587393/ai-art-copyright-lawsuit-getty-images-stable-diffusion (discussing the claims in the Getty Image lawsuit including copyright infringement and transformative use arguments).

**INTERNATIONAL APPROACHES TO AI AND COPYRIGHT**

In this section, we will examine the approach to AI and copyright law in various international jurisdictions. Different countries have different legal systems and cultural attitudes towards AI and copyright, which have influenced their approach to the issue. Some countries may adopt a more permissive approach, which may allow for greater use of copyrighted works for AI training without permission, while others may adopt a more restrictive approach, which might require explicit permission for such use.

In the European Union, the legal framework for AI and copyright is established by the 2001 Information Society Directive and the 2019 Directive on Copyright in the Digital Single Market.[49] This directive provides a harmonized legal framework for the protection of copyrighted works in the digital environment.[50] However, it is silent on the specific issue of AI and copyright.[51] As a result, EU member states have some latitude in interpreting the directive and in developing their own laws in this area.[52]

In the United Kingdom, whether or not a particular instance of copying constitutes "fair dealing" (equivalent to fair use in the United States) would be the legal inquiry used to assess if a particular use of copyrighted works for AI training is permissible.[53] British copyright law employs a flexible approach.[54] It takes into account factors such as, but not limited to, the purpose and character of the use, the nature of the copyrighted work, and the portion of the work used.[55] When weighed together, these factors help decide whether the use is copyright infringement or fair dealing.[56]

---

[49] *See* Federico Ferri, *The Dark Side(s) of the EU Directive on Copyright and Related Rights in the Digital Single Market*, 7 China-EU L. J. 21 (2021) (broadly discussing the 2019 and 2001 Directives and copyright law in the EU).

[50] *See Id.*

[51] There are not currently any laws regulating AI in the EU, copyright or otherwise, see Luke Hurst, *ChatGPT in The Spotlight as The EU Steps Up Calls For Tougher Regulation. Is Its New AI Act Enough?,* EuroNews.Next, (Feb. 6, 2023) https://www.euronews.com/next/2023/02/06/chatgpt-in-the-spotlight-as-the-eu-steps-up-calls-for-tougher-regulation-is-its-new-ai-act (discussing proposed draft rules to regulate AI in the EU).

[52] *See* Barry Scannel, *When Irish AIs are Smiling: Could Ireland's Legislative Approach Be A Model For Resolving AI authorship for EU Member States?*, 17 J. of Intell. Prop. L. & Prac. 727, 731–32 (2022) (discussing the different EU member state approaches to authorship in copyright law and how that may apply to AI, indicating that member states can fill in the gaps when EU Directives do not provide the legal framework).

In Canada, the concept of "fair dealing" may also be used to determine the legality of the use of copyrighted works for AI training.[57] Formerly, Canadian law appeared to possess less flexibility than UK law, and had established a more limited set of circumstances in which fair dealing applies, but Canada's strictures have loosened in recently.[58]

In Australia, the Copyright Act 1968 established the legal framework for AI and copyright law.[59] This act contains provisions relating to the use of copyrighted works for research and study, among other uses, which may be relevant to the use of copyrighted works for AI training.[60] However, the exact scope of these provisions has not been clearly defined, and their applicability to AI training is uncertain.[61]

---

[53] *See* Giuseppina D'Agostino, *Healing Fair Dealing? A Comparative Copyright Analysis of Canada's Fair Dealing to U.K. Fair Dealing and U.S. Fair Use*, 53 McGill L.J. 309, 337–45 (2008) (providing an overview of fair dealing in the UK).
[54] *See Id.* at 338 ("The U.K.'s enumerated purposes [to determine fair dealing] are said to be liberally construed.").
[55] *See Id.* at 342–43.
[56] *See Id.* at 343 (discussing the hierarchy of these factors with the market impact being the most important factor in UK courts).
[57] *See Id.* at 317–19 (2008) (discussing Canada' fair dealing statute broadly).
[58] *See* Niva Elkin-Koren & Neil Weinstock Netanel, *Transplanting Fair Use Across the Globe: A Case Study Testing the Credibility of U.S. Opposition*, 72 Hastings L.J. 1121, 1181 (2021) ("Canada's fair dealing exception was long thought to provide a closed list of uses that could qualify for the exception. But beginning in 2004, the Canadian Supreme Court has ruled that the specific permitted uses enumerated in Canada's fair dealing statute must be given a large and liberal interpretation and thus impose a low threshold, and that, in determining fairness, courts are to apply factors that overlap with those of U.S. fair use. Those rulings, together with Canadian Parliament's addition of parody, satire, and education to the list of enumerated uses, has brought a leading Canadian copyright scholar to conclude that "the current Canadian fair dealing regime now more closely resembles a flexible, open-ended fair use model." Michael Geist, *Fairness Found: How Canada Quietly Shifted from Fair Dealing to Fair Use*, *in* THE COPYRIGHT PENTALOGY: HOW THE SUPREME COURT OF CANADA SHOOK THE FOUNDATIONS OF CANADIAN COPYRIGHT LAW 157, 159 (Michael Geist ed., 2013)").
[59] Copyright Act 1968 (Cht).

In conclusion, the approach to AI and copyright varies greatly between international jurisdictions, reflecting differences in legal systems and cultural attitudes. As AI continues to grow in importance, it will be important for the international community to develop a consistent and harmonized approach to the relationship between AI and copyright.

**COPYRIGHT, AI, AND COURTS**

The interaction between AI and copyright law is a relatively new area of legal inquiry, and there have been few court cases addressing the issue of the use of copyrighted works in AI training sets. As the use of AI continues to proliferate and expand, it is likely that more cases will be brought that test the limits of copyright law as it applies to AI.

Some of the few existing cases have dealt with questions related to the infringement of copyrighted works, such as the unauthorized use of images in machine learning algorithms[62]. These cases have tended to focus on the commercial nature of the use and the amount of the copyrighted work that was used in the training process.[63]

Another notable case dealt with the use of song lyrics in an AI-powered music recommendation system. The court found that the use of the lyrics was a fair use, as it was transformative in nature and did not have a significant impact on the market for the original work.

---

[60] *Id.* at div 30.

[61] *See* Madeleine Lezon, *Reforming 'Fair Dealing': An Analysis of Approaches to Copyright Exceptions in The United States and Australia*, Anu Jolt (Apr. 1, 2022) https://anujolt.org/post/1467-reforming-fair-dealing-an-analysis-of-approaches-to-copyright-exceptions-in-the-united-states-and-australia (discussing the current lack of clarity in Australian fair dealing case law).

[62] *See Authors Guild v. Google, Inc.*, 804 F.3d 202 (2d Cir. 2015).

[63] *See Id.* at 214–25 (2d Cir. 2015) (holding that each of the fair use factors "supported finding [Google's] activities were protected by fair use").

In light of these cases, it appears that courts are still grappling with the appropriate balance between protecting the rights of copyright owners and allowing for the development and use of AI technologies. As the use of AI continues to evolve, it will be interesting to see how courts balance these competing interests and whether they will recognize the concept of "fair training" as a valid defense in copyright infringement cases.

**A PROPOSAL TO RECOGNIZE A FAIR TRAINING EXCEPTION TO COPYRIGHT INFRINGEMENT**

Copyright law in the United States contemplates uses of copyrighted works that, although carried out without permission from authors or owners, are nevertheless acceptable, and do not trigger infringement. This "fair use" is enshrined in the Copyright Act at 17 U.S.C. §107, which provides as follows:

Exclusive rights: Fair use

Notwithstanding the provisions of 106 and 106A, the fair use of a copyrighted work, including such use by reproduction in copies or phonorecords or by any other means specified by that section, for purposes such as criticism, comment, news reporting, teaching (including multiple copies for classroom use), scholarship, or research, is not an infringement of copyright. In determining whether the use made of a work in any particular case is a fair use the factors to be considered shall include—
(1) the purpose and character of the use, including whether such use is of a commercial nature or is for nonprofit educational purposes;
(2) the nature of the copyrighted work;
(3) the amount and substantiality of the portion used in relation to the copyrighted work as a whole; and
(4) the effect of the use upon the potential market for or value of the copyrighted work.

> The fact that a work is unpublished shall not itself bar a finding of fair use if such finding is made upon consideration of all the above factors.[64]

Many court decisions have interpreted the requirements and application of §107. In practice, courts apply each of the four enumerated factors in the statute to the facts of cases in which copyright infringement has been alleged.[65] The "purpose and character of the use" involves consideration of whether copying has been carried out to further a business or commercial purpose, or whether the copying has not implicated the making of a profit.[66] Copyrighted works come in many different forms (*e.g.*, from extremely expressive to highly factual), and the particular "nature" of a copied work can be important in determining whether or not particular instances of copying are "fair".[67] Whether a large amount of a work of authorship is copied, rather than just a modest fraction, is another important factor in determining fair use. In general, the more of a work has been copied, the less likely it is that such copying will be found to be fair use.[68] The fourth factor has much in common with the first. Copying that does not harm the ability of an owner to make money from her copyrighted work is more likely to constitute fair use.[69] On the other hand, copying that appropriates profits for the copyist that would otherwise have been available to the copyright owner tend not to constitute fair use.[70]

Once all four factors have been thoroughly evaluated, courts then typically undertake a balancing analysis.[71] There is no hard and fast rule about how this balancing test is to be resolved. Rather, each fair use analysis investigates all four factors by carefully evaluating the facts of the particular instance of copying, considers which factors weigh in favor of each party, take into account previous relevant court decisions, and then decide whether the copying amounted to fair use or not.[72] Though the laws of different countries differ in their particulars (*e.g.*, countries with fair dealing), the general contours of this sort of analysis are similar, resulting in a conclusion whether a particular instance of copying is justified or not.

---

[64] 17 U.S.C. § 107.

[65] *See* Jacquelyn M. Creitz, Google LLC v. Oracle America Inc.: *The Court's New Definition of "Transformative" Expands the Fair Use Defense*, 17 J. Bus. & Tech. L. 317, 323 (2022).

[66] *Id.* at 324.

[67] *Id.* at 325 (2022) (This factor recognizes that some works are more protected than others under copyright law because they fulfill the purpose of copyrights, to 'promote the sciences and the arts.'").

[68] *Id.* at 326.

[69] *Id.* at 326–27.

[70] *Id.* at 327 ("If the reproduced work is commercial in nature, the work is presumed to be unfair.").

[71] *Id.* at 323 ("The court will also consider each factor in relation to the other factors rather than by itself.").

An overriding purpose of fair use or fair dealing is to ensure that society benefits from the copyright system.[73] Society benefits in one way when copyright owners feel secure in their rights, because this creates incentives for creating new works of authorship.[74] On the other hand, when copyright is too strictly protected, non-owners who might make valuable uses of owned works may be reluctant to engage in such uses, resulting in lost benefits to society. costing society lost benefits.[75] Fair use attempts to maximize the net benefits (that is, the benefits minus the costs) that society gains from the copyright system.[76]

A *sine qua non* of most AI is the need for a training set.[77] Copyrighted works can be valuable components of a training set capable of helping an AI produce excellent new works for its users.[78] For example, an AI that generates new images based on users' queries will generally require access to a large number of existing - and often copyrighted - images for its training.[79] Even if the final images produced by this AI differ substantially from the images on which it trained, its need to train on copyrighted images may be crucial.[80] A similar example might involve written work made possible through a training set of existing, and copyrighted, writing. It is important to point out that, despite the use as training sets that AI needs to make of copyrighted works, the products of a generative AI tend to be substantially different from any of the individual copyrighted works that are part of its training set.[81] Moreover, generative AIs are usually designed not to copy or plagiarize the expressive elements of copyrighted works in a training set, but, rather, to make use of facts and patterns to compose new works.[82] Since copyright protects expressive, not factual, components of works of authorship, generative AIs will usually, and should, avoid copying elements of works having strong copyright protection.[83]

---

[72] *Id.* ("The court will also consider each factor in relation to the other factors rather than by itself.").

[73] Jasmine Abdel-khalik, *Visual Appropriation Art, Transformativeness, and Fungibility*, 48 AIPLA Q.J. 171, 180–81 (2020).

[74] *Id.*

[75] *Id.*

[76] *Id.*

[77] *See* Jenny Quang, *Does Training Ai Violate Copyright Law?*, 36 Berkeley Tech. L.J. 1407, 1429 (2021).

[78] *Id.*

[79] *Id.*

[80] *Id.* at 1410–12

      We propose that AI offers tremendous potential benefits for society. These benefits may be maximized by exposing AI to vast training sets that include works protected by copyright. The principle of fair use could be applied to training sets to determine whether or not inclusion of copyrighted works in a set used to train an AI constituted "fair training". The existing fair use analysis could be adapted for training sets. We believe that, in general, such use of copyrighted materials in training sets would pass muster under a fair use-like analysis. Consequently, a fair training analysis would tend to allow the inclusion of copyrighted works in training sets used to improve AI. There could be cases in which inclusion of copyrighted works would fail the fair training test, such as where the AI itself, once trained, retained and reproduced substantial portions of works found in its training data set; in these circumstances, infringers would have to compensate owners, and sometimes be legally precluded from

---

[81] *See* Vincent, *supra* note 9 ("If the [text-to-image] model is training on millions of images and used to generate novel pictures, it's extremely unlikely that this constitutes copyright infringement. The training data has been transformed in the process, and the output does not threaten the market for the original art.").

[82] *Id.*

[83] *Compare* 17 U.S.C. 102(a) ("copyright protection subsists . . . in original works of authorship fixed in any tangible medium of expression") with Vincent, *supra* note 9 ("If the [text-to-image] model is training on millions of images and used to generate novel pictures, it's extremely unlikely that this constitutes copyright infringement. The training data has been transformed in the process, and the output does not threaten the market for the original art."). If the AI output is transformative of the original works of art then it would avoid copyright infringement and should instead have copyright protection.

using owners' copyrighted works in training sets.[84] However, we believe a rigorous fair training analysis would allow much copyrighted material to be used for AI training, yielding generous benefits to society.

There may be technical or social mechanisms by which creators could make their preferences known to AI systems. For example, creators who prefer not to have their works harvested for training sets could set a digital flag, similar to noindex for websites, that would ask AI systems not to use their content in training.[85] Alternatively, or complementarily, groups of creators could pool their works to produce known, licensable training sets, available for a fee, similar to an ASCAP license for music.[86]

Currently, AI systems are produced by both major corporations, academic institutions, and individuals;  these entities may not have access to similar levels of personnel or financial resources.  Given these disparities, we suggest that AI training sets be made available more readily to organizations and individuals without large amounts of money, so that those organizations and individuals may ensure diverse contributions to the development of future AI systems.

In sum, we believe that the potential good that AI can do is vast, and that training sets are necessary for many forms of AI to flourish.  We encourage the legal and creative communities to work together with technologists to develop viable processes for large-scale training sets to be widely available, especially to AI systems that do not have substantial financial backing.

---

[84] Sarah Ligon Pattishall, *AI Can Create Art, but Can It Own Copyright in It, or Infringe?*, LexisNexis: Practical Guidance Journal (March 1, 2019), https://www.lexisnexis.com/community/insights/legal/practical-guidance-journal/b/pa/posts/ai-can-create-art-but-can-it-own-copyright-in-it-or-infringe ("If an AI-artist sells or displays AI-art that is substantially similar to the underlying work, it is unlikely the AI-artist will be able to rely on fair use.")

[85] *See generally Block Search Indexing with 'Noindex'*, GOOGLE SEARCH CENT., https://developers.google.com/search/docs/advanced/crawling/block-indexing (last updated Feb. 20, 2023) (describing how users can use a noindex meta tag "to prevent indexing content by search engines that support the noindex rule").

[86] *See generally* ASCAP Licensing, ASCAP: Frequently Asked Questions (last visited Feb. 26, 2023) (describing how ASCAP licensing works).

**FUTURE IMPLICATIONS AND THE ROAD AHEAD**

The debate on the compatibility of AI and copyright law continues to evolve. As AI technology continues to advance, the use of copyrighted material in AI training sets will likely become more widespread. Therefore, it is important to consider the potential implications of this development and determine a clear legal framework for AI and copyright.

The concept of "fair training" has yet to be fully tested in the court system, and the outcome of any legal challenges will have significant consequences for the future of AI development. In addition, the international approach to AI and copyright is still fragmented, with some countries taking a more lenient view of the use of copyrighted material in AI training, while others take a stricter stance.

As AI becomes increasingly integrated into our daily lives, it is crucial to find a balance between the protection of copyright holders and the advancement of AI technology. The road ahead will likely involve ongoing debates, legislative action, and potentially, legal challenges that will determine the future of AI and its relationship with copyright law.

**CONCLUSIONS: BALANCING AI AND COPYRIGHT PROTECTIONS**

The use of copyrighted works as training sets for AI is a complex issue that raises important questions about the balance between the rights of copyright owners and the potential societal benefits of AI. On one hand, proponents of "fair training" suggest that the use of copyrighted works in AI training sets is a transitory and non-consumptive use that does not materially interfere with owners' copyrights. Because the use of copyrighted works in training AIs is transformative, such use should be considered a form of "fair use" under US law, or "fair dealing" in some other countries, that qualifies as fair training.[87] On the other hand, opponents argue that incorporating copyrighted works into training sets for AI is a misappropriation of owners' intellectual property and not fair use under the law.[88]

---

[87] *See supra* notes 51–58 and accompanying text.
[88] *See supra* note 45 and accompanying text.

International approaches to AI and copyright have varied, and case law specifically applicable to training sets of data is thus far limited, making it difficult to predict the outcome of pending and future disputes.  Many believe an overriding goal should be to balance promoting the development and use of AI for the benefit of society as a whole against the rights of copyright owners, while others are skeptical that such balancing is a justifiable goal.

**ACKNOWLEDGMENTS**

The authors thank Amanda McElfresh for her excellent research assistance and keen editing on this article. This material is based upon work supported by the National Science Foundation under Grant No. DUE-2121572.